\PassOptionsToPackage{dvipsnames}{xcolor}
\documentclass[11pt,letterpaper]{article}
\usepackage[margin=1in]{geometry}
\usepackage{setspace}

\usepackage{tikz}
\usetikzlibrary{arrows,arrows.meta}
\usepackage{filecontents}

\usepackage{mathtools}

\usepackage{float}
\usepackage{caption, subcaption}
\usepackage{graphicx}
\usepackage{amsmath,amsthm,amssymb,amsfonts}
\usepackage{algorithm}
\usepackage{listings}
\usepackage[colorlinks,linkcolor=blue]{hyperref}
\usepackage{url}
\usepackage{cleveref}

\usepackage{colortbl}

\usepackage{pdfpages}

\newtheorem{definition}{Definition}
\newtheorem{remark}{Remark}

\newcommand{\hlcolor}{Yellow!35}
\newcommand{\hlcolorTwo}{LimeGreen!35}
\makeatletter
\newenvironment{btHighlight}[1][]
{\begingroup\tikzset{bt@Highlight@par/.style={#1}}\begin{lrbox}{\@tempboxa}}
{\end{lrbox}\bt@HL@box[bt@Highlight@par]{\@tempboxa}\endgroup}

\newcommand\btHL[1][]{%
  \begin{btHighlight}[#1]\bgroup\aftergroup\bt@HL@endenv%
}
\def\bt@HL@endenv{%
  \end{btHighlight}%
  \egroup
}
\newcommand{\bt@HL@box}[2][]{%
  \tikz[#1]{%
    \pgfpathrectangle{\pgfpoint{1pt}{0pt}}{\pgfpoint{\wd #2}{\ht #2}}%
    \pgfusepath{use as bounding box}%
    \node[anchor=base west, fill=\hlcolor,outer sep=0pt,inner xsep=1pt, inner ysep=0pt, rounded corners=2pt, minimum height=\ht\strutbox+2pt,#1]{\raisebox{1pt}{\strut}\strut\usebox{#2}};
  }%
}

\newenvironment{btHighlightTwo}[1][]
{\begingroup\tikzset{bt@HighlightTwo@par/.style={#1}}\begin{lrbox}{\@tempboxa}}
{\end{lrbox}\bt@HLTwo@box[bt@HighlightTwo@par]{\@tempboxa}\endgroup}

\newcommand\btHLTwo[1][]{%
  \begin{btHighlightTwo}[#1]\bgroup\aftergroup\bt@HLTwo@endenv%
}
\def\bt@HLTwo@endenv{%
  \end{btHighlightTwo}%
  \egroup
}

\newcommand{\bt@HLTwo@box}[2][]{%
  \tikz[#1]{%
    \pgfpathrectangle{\pgfpoint{1pt}{0pt}}{\pgfpoint{\wd #2}{\ht #2}}%
    \pgfusepath{use as bounding box}%
    \node[anchor=base west, fill=\hlcolorTwo,outer sep=0pt,inner xsep=1pt, inner ysep=0pt, rounded corners=2pt, minimum height=\ht\strutbox+2pt,#1]{\raisebox{1pt}{\strut}\strut\usebox{#2}};
  }%
}

\newcolumntype{L}[1]{>{\raggedright\let\newline\\\arraybackslash\hspace{0pt}}m{#1}}
\newcolumntype{C}[1]{>{\centering\let\newline\\\arraybackslash\hspace{0pt}}m{#1}}
\newcolumntype{R}[1]{>{\raggedleft\let\newline\\\arraybackslash\hspace{0pt}}m{#1}}

\author{Anonymous author}

\title{Model-based Testing of Practical Distributed Systems in Actor Model}
\date{
$^1$ITMO University\\
$^2$VK.com
}

\begin{document}

%
%
%
\author{Ilya Kokorin$^{1,2}$ \and
Evgeny Chernatskiy$^{1,2}$ \and
Vitaly Aksenov$^1$}

%
%
%
\maketitle              
\begin{abstract}
Designing and implementing distributed systems correctly can be quite challenging. Although these systems are often accompanied by formal specifications that are verified using model-checking techniques, a gap still exists between the implementation and its formal specification: there is no guarantee that the implementation is free of bugs.

To bridge this gap, we can use model-based testing. Specifically, if the model of the system can be interpreted as a finite-state automaton, we can generate an exhaustive test suite for the implementation that covers all possible states and transitions.

In this paper, we discuss how to efficiently generate such a test suite for distributed systems written in the actor model. Importantly, our approach does not require any modifications to the code or interfering with the distributed system execution environment. As an example, we verified an implementation of a replication algorithm based on Viewstamped Replication, which is used in a real-world system.

\vspace{0.5cm}

\end{abstract}

\newpage

\section{Introduction}

Designing and implementing distributed systems correctly is extremely challenging, as real-world distributed systems have an enormous number of possible states, corner cases, potential failure scenarios, etc. Therefore, we cannot be certain that these implementations are correct, and we need a comprehensive method to test them. Basing our distributed systems on existing verified algorithms does not solve all our problems. Indeed, a practical distributed system typically requires additional features beyond the basic algorithms. 

As an example, we use a replication algorithm developed in VK.com based on the mature well-tested Viewstamped Replication algorithm~\cite{liskov2012viewstamped,oki1988viewstamped}. Although the algorithm from the paper is well-described and even formally verified~\cite{vrtla}, it had to be modified in a couple of ways to satisfy the production needs:


\begin{itemize}
    \itemsep=0em
    \item In the original Viewstamped Replication algorithm, nodes sometimes send the whole log to other nodes. This is impossible in real-world scenarios since the database log can be up to hundreds of terabytes in size. To solve this, the new algorithm supports incremental downloading: it sends only the missing parts of the log without sending the parts that the other node already has.

    \item To reduce the load, it was required to split a highly-loaded database shard, consisting of $N$ replicas $R_1, R_2, \ldots R_n$, synchronized with their replication algorithm, into $M$ distinct sub-shards, with the first sub-shard consisting of $N$ replicas $R_1^1, R_2^1, \ldots R_n^1$, the second sub-shard consisting of $N$ replicas $R_1^2, R_2^2, \ldots R_n^2$, and so on.

    \item Also, the replicas are required to periodically do garbage collection resembling the vacuum in PostgreSQL~\cite{shaik2020execute}, file compaction from Cassandra~\cite{lakshman2010cassandra} or Bigtable~\cite{chang2008bigtable}. Since the garbage collection procedure consumes a significant amount of disk, memory, and CPU resources, thereby, severely limiting the replica's ability to answer queries quickly, the new replication algorithm must ensure that no more than $M$ replicas out of $N$ perform garbage collection simultaneously. The original Viewstamped Replication algorithm is unaware of such a requirement, so it had to be modified.
\end{itemize}

The industry further convinces us that it is ubiquitous to modify algorithm from the paper for the production needs, e.g., Megastore~\cite{baker2011megastore} presented a modification of the Paxos~\cite{lamport2001paxos,lamport2019part} algorithm for speeding up replication process; CockroachDB~\cite{taft2020cockroachdb} authors implemented custom transaction processing algorithm, not similar to well-known and proven Two-Phase Commit~\cite{atif2009analysis}, Three-Phase Commit~\cite{atif2009analysis} or Calvin~\cite{thomson2012calvin} transaction processing protocols; TiDB~\cite{huang2020tidb} presented a variant of Raft~\cite{ongaro2015raft} consensus algorithm with the support of shard splitting mentioned above; Percolator~\cite{peng2010large} uses a modified variant of Two-Phase Commit protocol for transaction processing; Spanner~\cite{corbett2013spanner} uses Two-Phase Commit in Paxos groups for transaction processing in replicated groups and augment these algorithms with advanced timestamp management system called TrueTime~\cite{brewer2017spanner}; WeChat presented PaxosStore~\cite{zheng2017paxosstore} storage service that modifies the Paxos algorithm to make it more fault-tolerant.

In the well-known paper about practical Paxos implementation~\cite{chandra2007paxos} for the Chubby lock service~\cite{burrows2006chubby}, Google engineers summarized these kinds of challenges. They stated that ``there are significant gaps between the description of the Paxos algorithm and the needs of a real-world system. In order to build a real-world system, an expert needs to use numerous ideas scattered in the literature and make several relatively small protocol extensions. The cumulative effort will be substantial and the final system will be based on an unproven protocol''. 

To guarantee that the protocol modification neither breaks any invariant of the initial protocol (e.g., that committed log entries are never rewritten in the Viewstamped Replication protocol) nor violates any of its own invariants (e.g., that never more than one replica is running the garbage collection procedure simultaneously) we have to test our practical protocol implementations thoroughly.
Sadly, verifying a distributed system is extremely challenging due to the internal non-determinism of the distributed execution: nodes can receive messages in arbitrary order, an unreliable network may drop or corrupt messages, any subset of nodes may stop functioning at any moment, etc.
This leads to an enormous number of possible execution scenarios and corner cases which may never be covered using an ordinary unit testing approach.
To cover more cases during the test of a complex system, it is advised to use fuzzing~\cite{liang2018fuzzing}.
However, it generates random execution scenarios and does not guarantee testing all possible execution scenarios, thus being less reliable than checking all the possible executions.

To address the needs of testing all possible states and corner cases of a distributed protocol, engineers use model checking approach: the distributed protocol along with its invariants is specified in a formal language, e.g., TLA+~\cite{lamport1999specifying} or Promela~\cite{neumann2014using}, and after that use a model checking tool similar to TLC~\cite{yu1999model} or SPIN~\cite{holzmann1997model} to check that all the desired invariants hold in all possible execution scenarios.

Although model checking can help us verify specifications of our protocols written in a specialized protocol definition language (e.g., TLA+ or Promela), it leaves a significant gap between the protocol specification and the system implementation. Indeed, real-world distributed systems cannot be implemented using either TLA+, Promela, or any other protocol definition language. Instead, developers use programming languages such as C++, Go, and Java, among others. Therefore, the protocol specification written in a formal language will significantly differ from the protocol implementation written in a commonly-used programming language: the implementation will have to deal with operating systems (network stack, persistent disks, sygnals, etc), resource allocation and reclamation, memory and CPU optimizations, and various other tasks that we are not taking into account when writing protocol specification.  

Therefore, we cannot be sure that our implementation is correct even if we have verified our specification using model checking tools similar to TLC. We address this problem by presenting an approach that generates an exhaustive test suite for an actor-based protocol implementation based on the output of a model checking tool. The test suite covers all possible execution paths of the distributed system within the given constraints and allows for the deterministic replay of erroneous execution paths for debugging purpose, while our approach allows generating the exhaustive test suite significantly faster than our counterparts. 

The work is structured as follows: in Section~\ref{existing-solutions-section}, we present an overview of existing solutions to this problem; in Section~\ref{overview-section}, we present an overview of our approach; in Section~\ref{graph-algorithm-section}, we present a novel algorithm for generating the smallest exhaustive test suite in polynomial time; in Section~\ref{practical-results-section}, we present results of applying our test system to real-world distributed systems in VK.com; finally, we conclude in Section~\ref{conclusion-section}. 

\section{Existing solutions}
\label{existing-solutions-section}

We highlight three tools capable of generating tests based on protocol specifications written in TLA+: Modelator~\cite{modelator}, Kayfabe~\cite{dorminey2020kayfabe}, and the tool by Wang et al.~\cite{wang2023model}.

Using Apalache~\cite{konnov2019tla+} model checker, Modelator~\cite{modelator} generates \textbf{random} execution scenarios that act as tests for the protocol implementation. This instrument has several major drawbacks: first, it can generate only random execution scenarios, whereas our goal is to test the system on all possible executions within certain constraints. Secondly, Apalache supports only a small subset of all TLA+ syntax features, whereas our goal is to provide full support for all TLA+ syntax features. Thirdly, Modelator can generate tests only for Python programs; Go and Rust programming languages were previously supported, but their support has been dropped. While our goal is to support at least Go and C++ programming languages, which are commonly used in VK.com to implement distributed protocols.

Kayfabe~\cite{dorminey2020kayfabe} uses the TLC model checker to generate the system state graph, whose vertices represent possible states of the system and whose edges represent possible actions of the system (e.g., receiving and processing a message sent by another node). After that, Kayfabe uses a Traveling Salesman Problem solver to generate a set of paths that visit each vertex of the graph at least once. This set of paths becomes the test suite for the implementation. The tool also has several major drawbacks. First of all, Kayfabe does not guarantee visiting each edge (i.e., state transition) at least once; it guarantees only visiting each possible state at least once, whereas the goal of our algorithm is to verify each possible state transition of the protocol. Secondly, Kayfabe reduces the problem of generating the test suite to the Traveling Salesman Problem, which is an NP-hard problem~\cite{zambito2006traveling}. Therefore, the tool requires either spending exponential time to solve the problem or using approximate solvers, thus sacrificing test suite quality. Our solution uses the exact polynomial algorithm to generate the minimal exhaustive test suite. Thirdly, Kayfabe does not support state transitions with arguments, i.e., it does not preserve the information about which node has received a particular message, which is strictly necessary for checking complex behavior. In contrast, our solution directly supports most of TLA+ syntax features, including arguments of state transitions. 

The tool by Wang et al.~\cite{wang2023model} also has several issues with solving our problem. Firstly, it uses a heuristic called Partial Order Reduction to reduce the number of tests by cutting uninteresting edges in the state graph, thus sacrificing the exhaustiveness of generated test suite. Secondly, the authors report that running a single test takes a couple of seconds while our solution runs thousands of tests per second per core. Lastly, just like Kayfabe, it does not support state transitions with arguments, i.e., it does not preserve the information about which node has received a particular message.

\section{Overview of the approach}
\label{overview-section}

\paragraph{High level overview. } Before delving into the details we provide a brief overview of our approach. Suppose we are given a distributed system written in the actor model (Section~\ref{sec:actors}). An actor is an entity that has a state and reacts to certain events, such as incoming messages, and can create new events, such as outcoming messages. These distributed systems enable the reproducible testing (Section~\ref{sec:emulation}) with a test being an ordered list of events. Given a test, the emulator takes events one by one and asks the corresponding actors to process them.

Now, we need to two questions: how to generate tests and how to verify correctness.
To answer the first question (Section~\ref{sec:tests}), we write a model of the system in TLA+ and, then, ask TLC to generate a transition graph. Since we want to test all the transitions of the implemented system, we cover all edges of this graph with paths starting at the initial state using a novel efficient algorithm (Section~\ref{graph-algorithm-section}). These paths are represented by a list of events, which can be used as tests.

To answer the second question (Section~\ref{sec:correctness}), we augment each implemented actor with a function \texttt{ToModel()} that transforms its state to the state of this actor in the TLA+ model. Then, when emulating a test, in each step, we check that the transformed states of all implemented actors are equal to the states of their model versions.

\subsection{Actor model}
\label{sec:actors}

Testing an arbitrary practical distributed system implementation is an extremely difficult task: indeed, at any moment, any node can send a message over the network, wait for a message over the network, try to persist some information on a disk, etc. Given that all nodes of a practical distributed system are working concurrently, the execution of such system becomes nondeterministic: indeed, the network can deliver messages in arbitrary orders with arbitrary delays, can drop or corrupt some messages, arbitrary nodes can suspend its execution (due to garbage collection, operation system swapping, etc) or fail at any time (possibly losing its transient in-memory state and keeping only persistent on-disk state), etc. Therefore testing distributed systems in real-world environment is almost impossible: we can neither guarantee that we have tested all the possible execution scenarios nor deterministically reproduce erroneous executions over and over for debug purpose.

To overcome this issue and enable exhaustive and deterministic testing of practical distributed systems, we propose implementing a distributed system in the actor model~\cite{hewitt2010actor}. An actor is an abstraction of a node in a distributed system. It maintains the state and can react to events occurring in the system. An event may, for example, be a message received from another node over the network or a response to a previously issued asynchronous write of some data.

While reacting to an event, the actor may change its state and produce an arbitrary number of requests for future events. For example, it can send a message to another node that will be later delivered as an event or call an asynchronous write of some data and later receive either an acknowledgement or an error.

We can illustrate this concept with an implementation of a simple actor that stores key-value pairs (Listing~\ref{actor-example-listing}) and receives two kinds of events:
\vspace{-0.2cm}

\begin{itemize}
    \itemsep=0em
    \item A message containing a request to find the value associated with a particular key and send this value to the requester (or \texttt{nil} if this key does not exist in the actor storage);
    \item A message containing a request to add a new key-value pair to the system and send the message about the update to all nodes of the system.
\end{itemize}
\vspace{-0.3cm}

\begin{lstlisting}[caption={Implementation of a simple actor for storing key-value pairs},escapeinside={(*}{*)},captionpos=t,numbers=none,label={actor-example-listing}]
Actor KeyValueStorage:
    // The actor state conststs of a simple key-value storage
    State: map<string, string> storage

    // Function that will react to the received message, modify actor state and 
    // send messages to another actors
    Behavior: list<OperationRequest> OnEvent(Event event):
        if event.type == GET_REQUEST:
            // Respond with a value corresponding to the requested key
            if storage.contains(event.key):
                value := storage[event.key]
                return [SendRequest(receiver=event.sender, message=ValueResponse(value))]
            else:
                return [SendRequest(receiver=event.sender, message=ValueResponse(nil))]
        elif event.type == SET_REQUEST:
            // Add new key-value pair to the storage and notify all other actors
            storage[event.key] = event.value
            messagesToSend := []
            for actor in System.Participants:
                messagesToSend.append(SendRequest(receiver=actor, message=KeyUpdated(key)))
            return messagesToSend
\end{lstlisting}

If a system can be designed as several communicating actors, it is said to be designed in the actor model. After the implementation of the actors, the system must provide the means to serve events, e.g., send and receive messages over the network, 
flush data to persistent disks asynchronously by actors requests and notify actors about the results of these flush requests.
The events may be of any complexity and depend only on what is expected from the system.

\begin{figure}[t]
     \centering
     \begin{subfigure}[b]{0.45\linewidth}
          \centering
          \includegraphics[width=\linewidth]{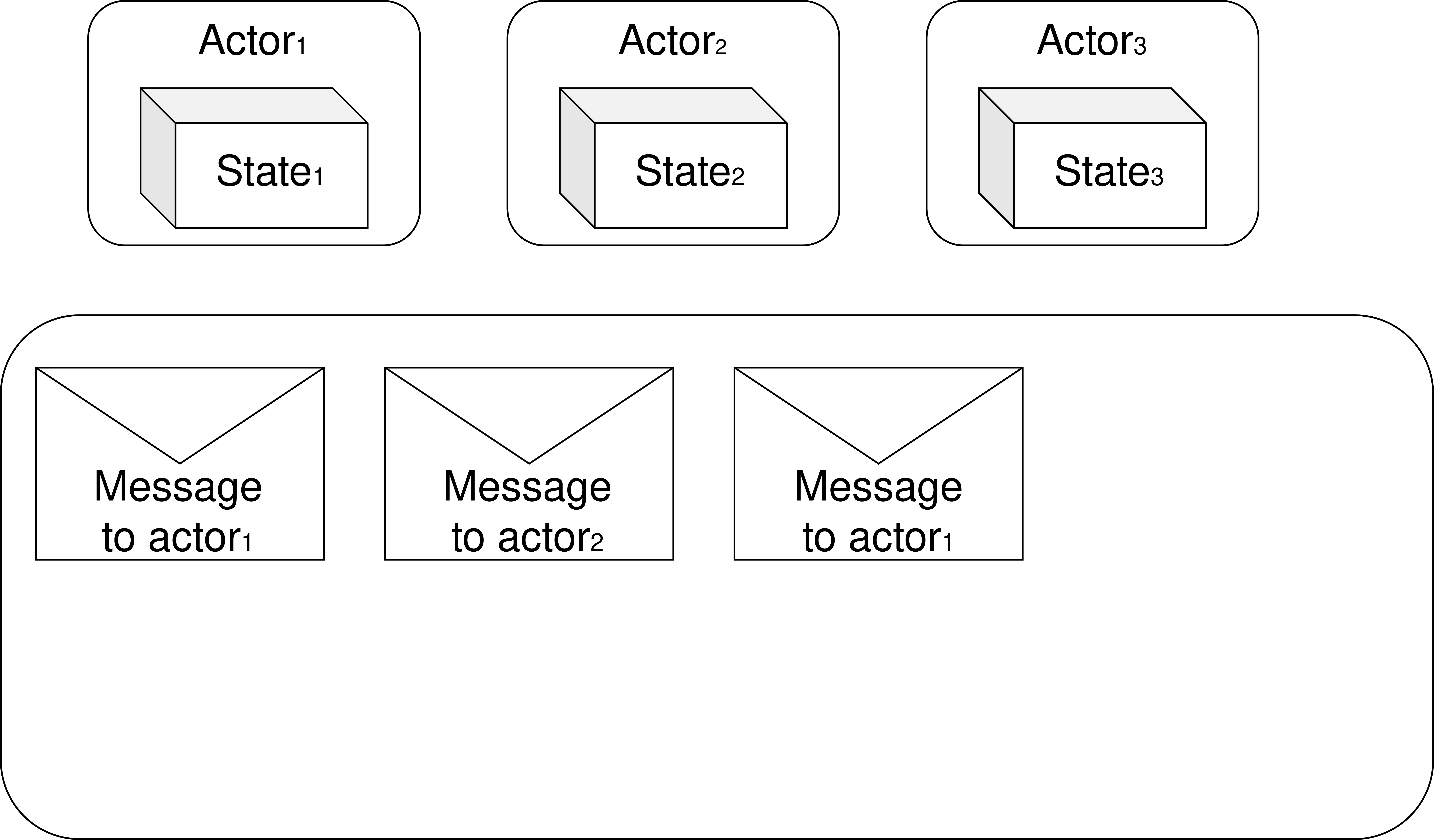}
          \caption{A collection of actors and a set of unprocessed events}
          \label{test-1-pic}
     \end{subfigure}
     \hfill
     \begin{subfigure}[b]{0.45\linewidth}
          \centering
          \includegraphics[width=\linewidth]{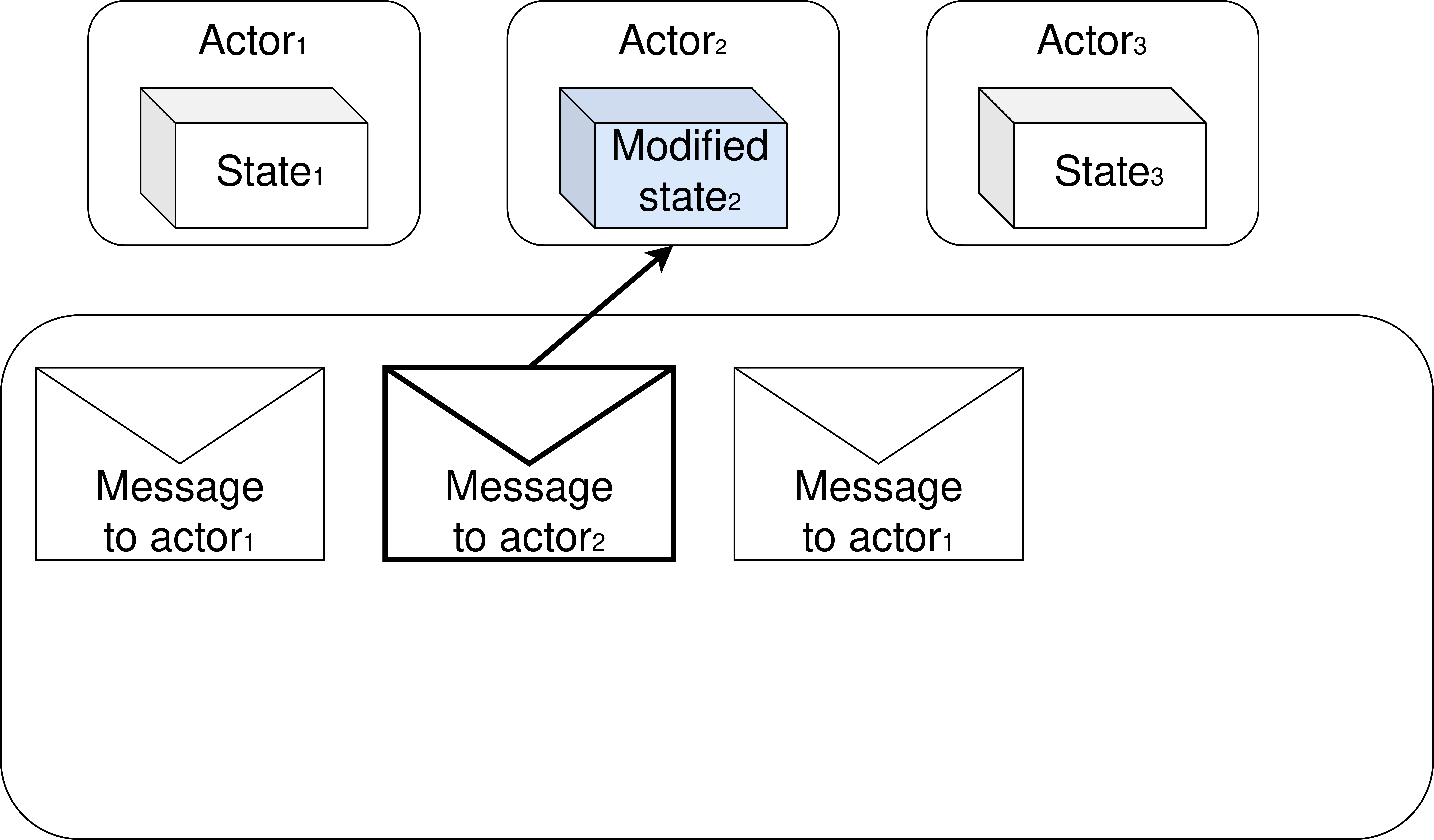}
          \caption{$Actor_2$ processes a chosen event from the unprocessed set modifying its state}
          \label{test-2-pic}
     \end{subfigure}
     \hfill
     \begin{subfigure}[b]{0.45\linewidth}
          \centering
          \includegraphics[width=\linewidth]{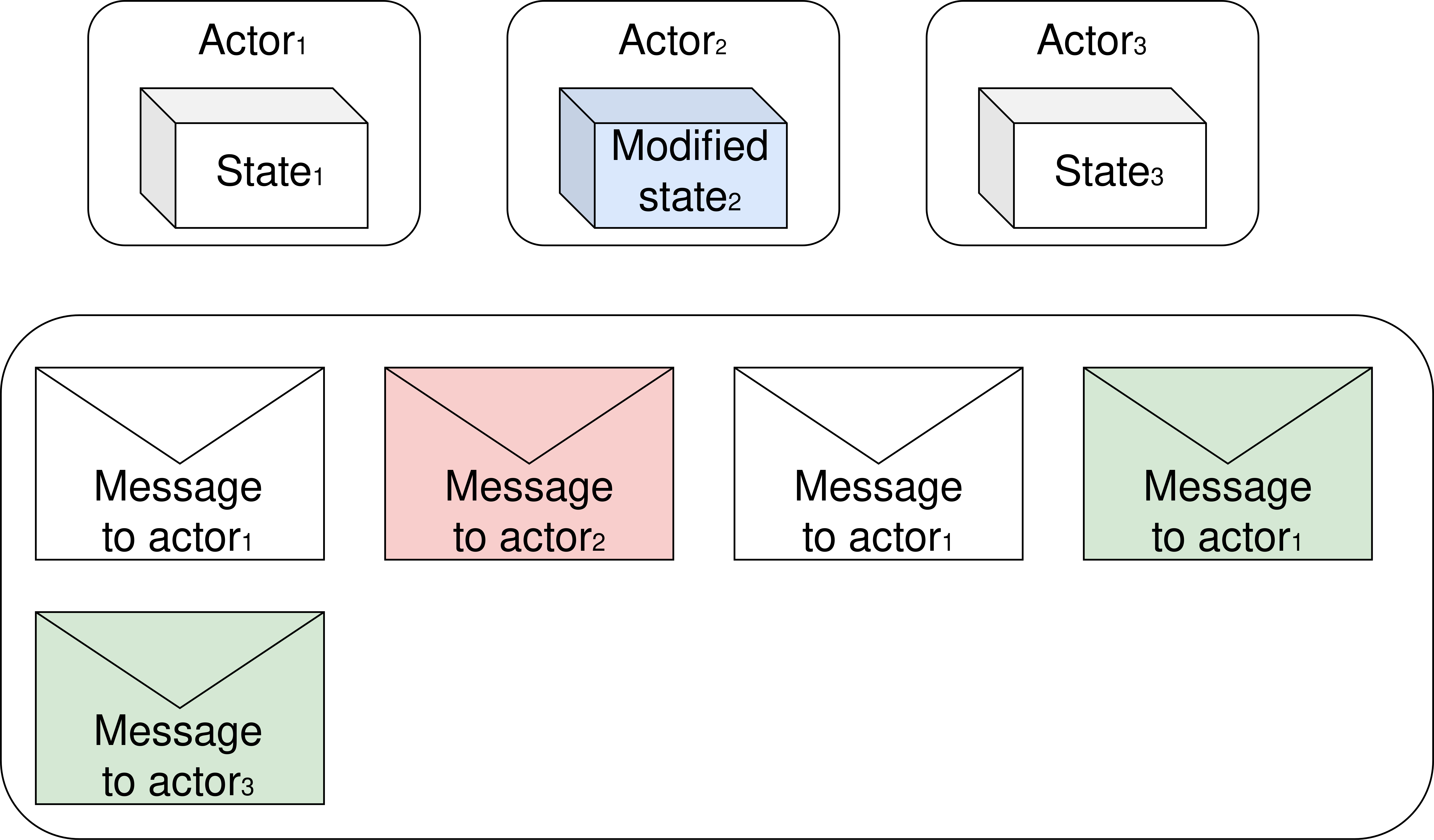}
          \caption{Remove the processed event from the set and add messages that were requested to send by $Actor_2$ to the unprocessed set}
          \label{test-3-pic}
     \end{subfigure}
     \hfill
     \begin{subfigure}[b]{0.45\linewidth}
          \centering
          \includegraphics[width=\linewidth]{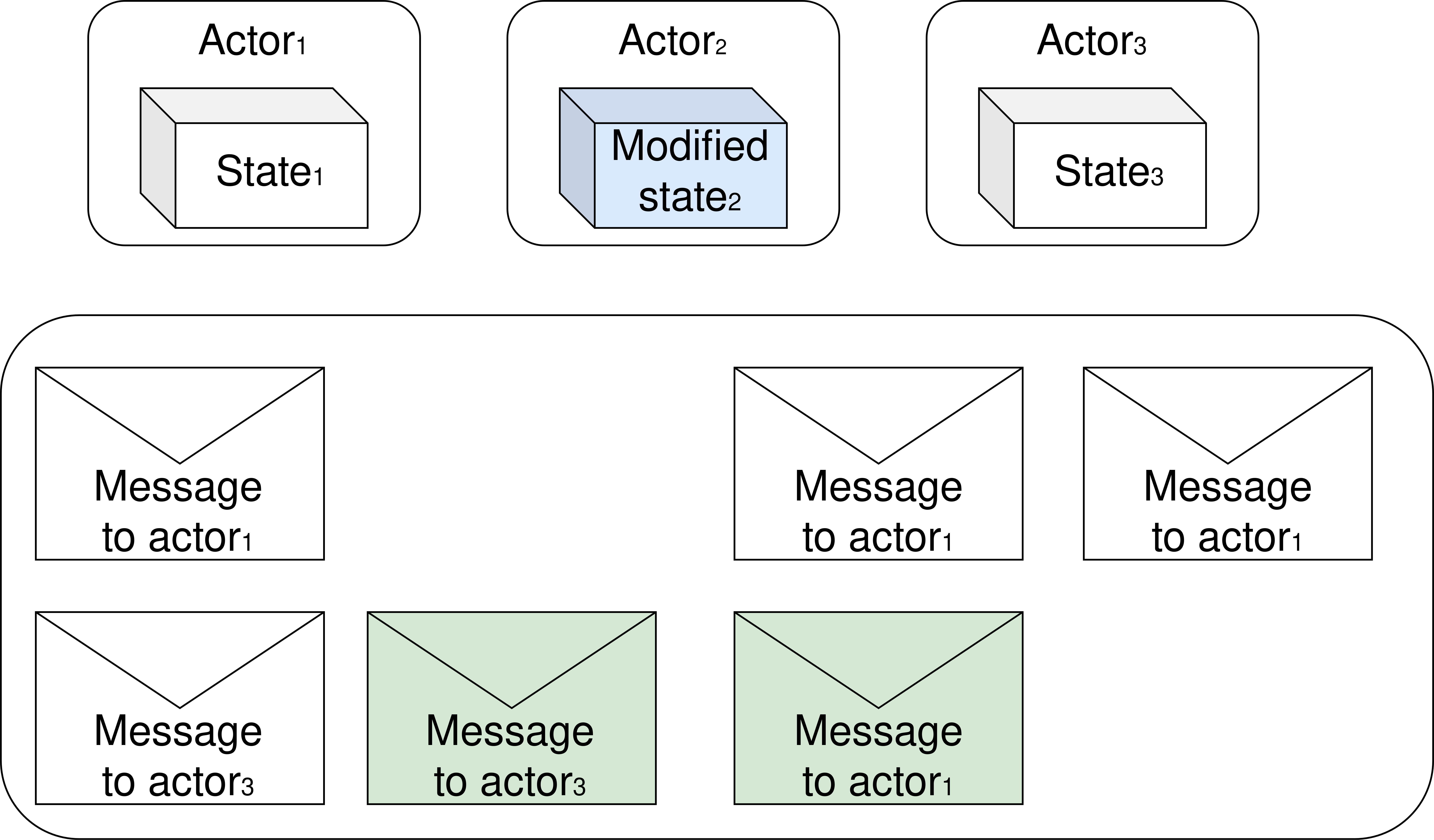}
          \caption{New events are added to the unprocessed set, simulating user requests from outside of the system under test}
          \label{test-4-pic}
     \end{subfigure}
    \caption{Testing actor system in a single-threaded environment}
    \label{test-pic}
\end{figure}

\subsection{Emulation of the system in the actor model for testing}
\label{sec:emulation}

Now, suppose we want to test an implementation of an arbitrary distributed system in the actor model. The simplest approach would be to run the stress test: we initialize actors on different machines and start generating random events. This approach has several downsides: for example, if the system fails, it will be difficult to recreate the scenario, and even if we can, debugging will be cumbersome.

Thus, for testing purposes, we will run the actor system in a single-threaded environment and will process all events synchronously, one by one (Fig.~\ref{test-pic}). This approach has the benefit of allowing deterministic replay of erroneous executions: whenever we find an erroneous execution path, we can deterministically replay it as many times as necessary to debug the implementation, since we know the exact order in which the system processed events. Within a single OS thread, we create all the actors that need to be tested (each actor is created with its initial state) and an initially empty set of unprocessed events. On each step, the testing algorithm:

\begin{enumerate}
    \itemsep=0em
    \item using programmed logic, withdraws a single event $e$ from a set of unprocessed events that should be processed by the corresponding actor $a$;

    \item calls event processing function of actor $a$ on event $e$ (Fig.~\ref{test-2-pic});

    \item collects operation requests returned by the event processing function and adds them to the set of unprocessed events (Fig.~\ref{test-3-pic});

    \item possibly adds several events (generated by some programmed logic) to the unprocessed set simulating messages from outside of our system, e.g., by an external client (Fig.~\ref{test-4-pic});

    \item proceeds to the next step, on which another unprocessed event may be delivered to another actor.
\end{enumerate}

Since all actors process events in a single thread and we know the order in which the events were processed, we can easily log any erroneous execution (if found) to a file and replay it later for debugging purposes. Such a testing method can use fuzzing~\cite{liang2018fuzzing} to generate interesting execution orders and events.

Using this approach, we can simulate different failures.
As the simplest ones, we can emulate the network failure by simply not delivering messages or corrupting them.
We may also emulate a crash and recovery of actors by splitting the state of each actor into ``persistent disk'' and ``volatile RAM'' parts. During testing, both parts will be stored in RAM. To simulate a crash, we clear the volatile part and stop delivering events to an actor and possibly drop some unprocessed events, e.g., data flushes and undelivered messages.

\begin{figure}[h]
  \centering
  \includegraphics[width=\linewidth]{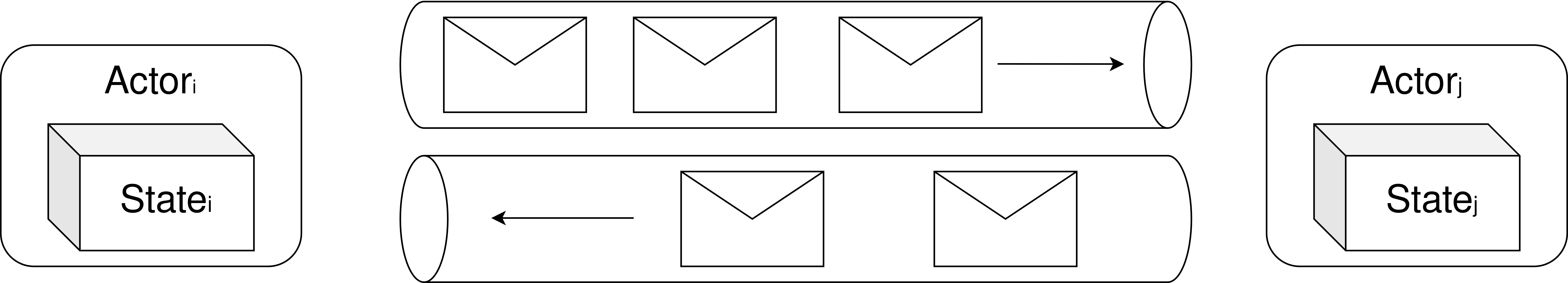}
  \caption{Modeling the set of unprocessed events as a collection of FIFO queues between each pair of actors} 
  \label{fifo-pic}
\end{figure}

Such a testing method assumes that events from the unprocessed set can be processed in any order. For example, messages in our system are transmitted using the UDP protocol and therefore can arrive and be processed by actors in any order. If our actors assumes some event delivery guarantee from the system (e.g., that between each pair of actors $A_i$ and $A_j$ messages are delivered in FIFO order) we should model the unprocessed events not as a plain set, but as a more complex structure, e.g. as a collection of FIFO queues of unprocessed messages between each pair of processes (Fig.~\ref{fifo-pic}) from which we may choose only the head elements for processing.

\subsection{Generation of tests}
\label{sec:tests}

In this section, we explain how to generate all tests with bounded parameters, which are then executed using the algorithm above and verified.

To check all the executions of fixed length, the developers use model checking. We implement a model of our distributed system in TLA+~\cite{lamport1999specifying}. Variables of the specification contain states of each actor and the collection of unprocessed events, which may be specified either as a simple unordered set or as a more complex data structure (e.g., a collection of pairs of FIFO queues)~--- in specifications of our production systems, we used a simple unordered set of unprocessed events. On initialization, each actor is set into an initial state, and the collection of unprocessed events is set to empty. For example, we may initialize the model of Viewstamped Replication algorithm the following way (Listing~\ref{tla-init-listing}): 

\begin{lstlisting}[caption={Possible intalization of a model correspondong to the implementation of Viewstamped Replication algorithm},escapeinside={(*}{*)},captionpos=t,numbers=none,label={tla-init-listing}]
Init ==
    // Each replica contains an initial state
    /\ replicaStates = [ replica \ in Replicas |-> [
        status |-> "Normal",
        log |-> <<>>,
        viewNumber |-> 0,
        commitNumber |-> 0,
        downloadReplica |-> replica,
        catchupPos |-> 0,
        phase2 |-> FALSE
    ]]
    /\ queriesCount = 0 // No user has issued a request to our system
    /\ unprocessedEvents = {} // The set of unprocessed events is initially empty
\end{lstlisting}

Transitions in the system occur through the following operations: adding a message from an external user to the set of unprocessed events, processing a single unprocessed event, and simulating a failure (e.g., corrupting an unprocessed message or clearing the volatile state of some actor). In Listing~\ref{tla-transition-listing}, we describe the transition corresponding to the start of the master election in Viewstamped Replication algorithm.

\begin{lstlisting}[caption={Possible description of a transition corresponding to the start of the master election in Viewstamped Replication algorithm},escapeinside={(*}{*)},captionpos=t,numbers=none,label={tla-transition-listing}]
TimeoutStartViewChange(replica) ==
    // The transition is possblie only if the number of master election does not 
    // exceed the threshould
    /\ ViewNumber(replica) < MaxViewNumber
    // Replica that started the master election process changes its state
    /\ replicaStates(*$'$*) = [replicaStates EXCEPT
        ![replica].status = "ViewChange",
        ![replica].viewNumber = @ + 1,
        ![replica].downloadReplica = "None",
        ![replica].catchupPos = 0.
        ![replica].phase2 = FALSE ]
    // Replica sends StartViewChange messages to all other replicas
    /\ unprocessedEvent(*$'$*) = unprocessedEvent (*$\cup$*) { [ 
            type |-> "StartViewChange",
            viewNumber |-> ViewNumber(replica) + 1,
            to |-> otherReplica
        ] : otherReplica \in Replicas }
    // The number of user requests does not change
    /\ UNCHANGED << queriesCount >> 
\end{lstlisting}

We also need to specify the invariants that algorithms must satisfy.
For example, for our algorithm we may specify the following ones (Listing~\ref{tla-invariants-listing}):

\begin{itemize}
    \item Prefix log consistency: for two replicas $R_i$ and $R_j$ committed prefixes of their logs must match exactly;

    \item Log commit progress: given that the number of master elections is limited, eventually the logs of all replicas become equal, and all log entries on each replica are committed.
\end{itemize}

\begin{lstlisting}[caption={Possible invariants of the Viewstamped Replication algorithm},escapeinside={(*}{*)},captionpos=t,numbers=none,label={tla-invariants-listing}]
PrefixLogConsistency ==
    \A r1, r2 \in Replicas:
        \/ IsPrefixOf(SubSeq(Log(r1), 1, CommitNumber(r1)),
                      SubSeq(Log(r2), 1, CommitNumber(r2)))
        \/ IsPrefixOf(SubSeq(Log(r2), 1, CommitNumber(r2)),
                      SubSeq(Log(r1), 1, CommitNumber(r1)))

LogCommitProgress ==
    []<> \A r \in Replicas:
            /\ Status(r) = "Normal"
            /\ LogQueryCount(r) = MaxLogQueryCount
            /\ CommitNumber(r) = LogLength(r)
\end{lstlisting}

After we wrote the whole specification with all the invariants, TLC model checker~\cite{yu1999model} builds a graph of all possible states and transitions of the system. Nodes of this graph are possible states of the distributed system (each system state consists of the actor states and the unprocessed events set), edges of this graph are transitions of the system state (a transition may be either processing of event by an actor, addition of new message to the unprocessed events set, crash and restart of an actor, etc). To make this graph finite and therefore allow TLC to build and check the graph, we provide an upper limit on certain system parameters, such as the number of user-provided messages and the number of master elections. If these limitations make the system graph finite, TLC builds it and checks that all the desired invariants hold for our model.

After that, we can be sure that the model is correct and satisfies all the desired invariants in the provided constraints. 
However, we cannot be sure in the correctness of the real-world implementation, since the transformation from a TLA+ specification to a commonly-used programming language is prone to bugs.

To verify whether our implementation aligns with the model, we must examine the graph generated by the TLC model checker. This graph is definitely not a tree because of the commutativity property~\cite{fischer1985impossibility}: if two messages $M_i$ and $M_j$ destined for two distinct processes $P_i \neq P_j$ exist in the set of unprocessed events at the same time at state $S$, then processing first $M_i$ then $M_j$ will lead to the very same state $S'$ as processing first $M_j$ then $M_i$. 


Now, we need to generate an \textit{exhaustive test suite} on which we will run the emulation from Section~\ref{sec:emulation}: a set of paths that cover each edge of the resulting graph at least once. Note that visiting each node of the graph at least once is not sufficient: indeed, we want to test the processing of each event in our system.

\begin{figure}[h]
  \centering
  \includegraphics[width=0.6\linewidth]{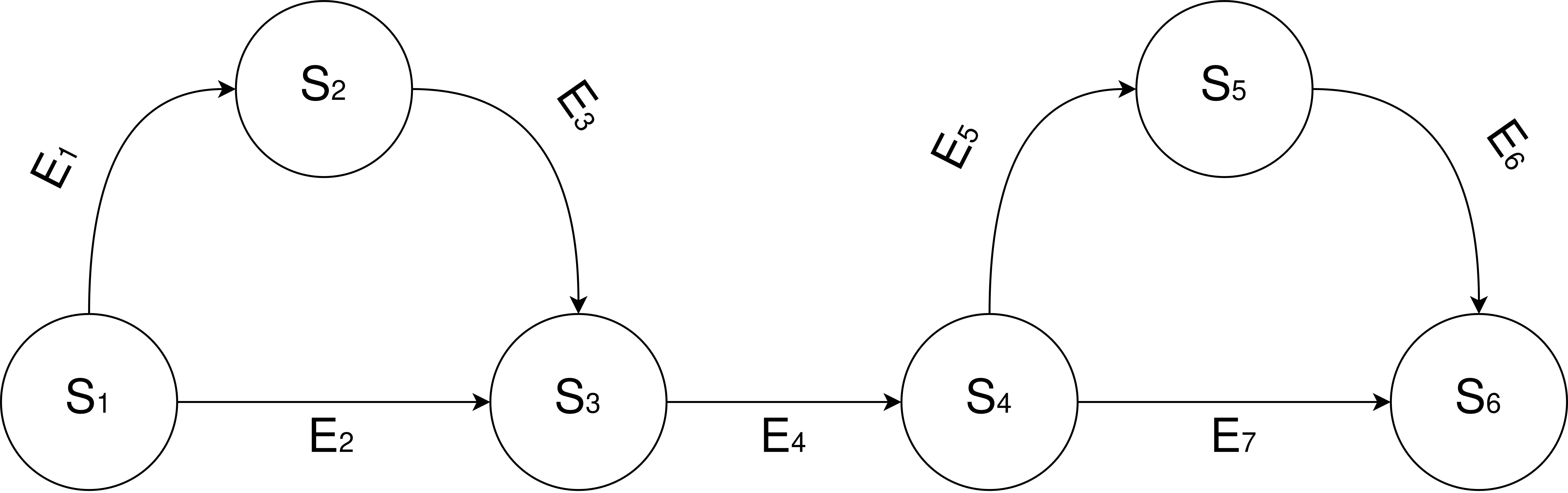}
  \caption{Example of graph that the TLC model checker can build} 
  \label{graph-example-pic}
\end{figure}

Consider the example of a graph in Fig.~\ref{graph-example-pic}. A single path $E_1, E_3, E_4, E_5, E_6$ visits all the nodes of the graph (i.e., all possible states of the distributed system) at least once. However, this test suite is not exhaustive. Indeed, we want to verify, for example, that processing event $E_2$ leads to the same system state as processing events $E_1$ and $E_3$. To make the test suite exhaustive, we can add path $E_2, E_3, E_7$. Moreover, we want this exhaustive test set to be as small as possible to reduce the testing time of our distributed system. In Section~\ref{graph-algorithm-section}, we present a novel, efficient algorithm for generating minimal exhaustive test suite.

\subsection{Correctness check}
\label{sec:correctness}

TLC model checker is able to not only build a transition graph for the TLA+ model and check all the desired invariants, but can also export the transition graph in \texttt{.dot} format used by \texttt{graphviz}~\cite{gansner2009drawing}.
However, this export does not preserve information about which event was executed by which actor during the transition.

To solve this issue, we designed a custom plugin for the TLC model checker. First, our plugin computes the set of covering paths over the transition graph using the algorithm from Section~\ref{graph-algorithm-section}.
%
%
Then, it dumps the information of all states (actors and set of unprocessed events) in the transition graph and after it dumps the computed paths:

\begin{itemize}
\item Each system state description consists of an identifier, states of all the actors in the system, and a set of unprocessed events. State of each actor is encoded as a set of variables, e.g., for the model of the Viewstamped Replication algorithm from Listing~\ref{tla-init-listing} the state of each actor contains: the status of the node (\texttt{``Normal''} or \texttt{``ViewChange''}); its stored log entries; its view number value; the number of committed entries in the log; the index of the actor from which the current actor downloads its log entries; the index of the currently downloaded log entry; and whether the current actor has seen a quorum of votes for the master of the current view. We assume that the state with the index \texttt{1} is the initial state of the algorithm.

\item Each path starts in the initial node \texttt{1}. For each path, we first store the number of edges in this path, and then the edges themselves. Each edge description consists of the operation corresponding to the edge (e.g., which message to add to the unprocessed event set, which event to process, which message to corrupt, which node to crash, etc) and the index of the destination system state. The description of each operation is provided in some machine-readable format (e.g., protobuf), not in some natural language.
\end{itemize}

The testing system must then ensure that the implementation is consistent with the model. The consistency is defined as follows: suppose we have a path $P = S_1 = S_{i_0} \rightarrow S_{i_1} \rightarrow S_{i_2} \rightarrow \ldots \rightarrow S_{i_N}$ from the graph built by TLC. The implementation is consistent with the model, if when consequently applying the operations on the path to the implemented actor system using the approach from Section~\ref{sec:emulation}, after each operation, the states of the implemented actor system coincide with the state of the model. In that case we can conclude that all the desired invariants hold for the implementation as well as for the model since the implementation states coincide with the model states on possible execution paths.






To verify whether the state of the implemented actor system corresponds to the model state, the user must implement the \texttt{ToModel()} method for each of its actors that should map the state of the implemented actor to the model variables.
For example, for our model from Listing~\ref{tla-init-listing}, \texttt{actor.ToModel()} method should return: the status of the actor (\texttt{``Normal''} or \texttt{``ViewChange''}); its stored log entries; its view number value; the number of committed entries in the actor log; the index of the actor from which the current actor downloads its log entries; the index of the currently downloaded log entry; and whether the current actor has seen a quorum of votes for the master of the current view.

The same \texttt{ToModel()} method should be implemented for all events so that the testing system can ensure that unprocessed events set in the actor system implementation correspond to the unprocessed events set from the model.

To execute a single test corresponding to a single path from the file with the exhaustive test suite generated by our TLC plugin the test system does the following:

\begin{enumerate}
    \itemsep=0em
    \item Initialize the actors of the system in the initial state and empty the unprocessed events set;

    \item Read the next edge from the current test path. The edge consists of the description of the action $A$ to perform and $i$~--- the index of the model state after applying this action;

    \item Apply the action $A$ to the implementation (e.g., add new message to the unprocessed event set, process an event from the unprocessed event set, corrupt a message, crash a node, etc);

    \item Fetch state $S_i$ with index $i$ from the file generated by the plugin;

    \item Call \texttt{ToModel()} method on each actor and ensure that the result of the call is equal to the state of the corresponding actor from the model state $S_i$. 

    \item Call \texttt{ToModel()} method on each event in the unprocessed events set and ensure that the unprocessed events set from the implementation corresponds to the unprocessed events set from the model state $S_i$;

    \item If there is an edge left, go to step (2), otherwise, finish the test execution.
\end{enumerate}

\section{Test suite generation algorithm}
\label{graph-algorithm-section}

In the previous section, we obtained a directed graph, known as the transition graph, where vertices represent the states of the system, and an edge between two states represents a transition from one state to another upon processing an event. Note that each edge has a corresponding an event and, thus, there may be several edges between two nodes initiated by different events. Now, our goal is to generate a test suite where each test is a sequence of events starting from the initial state, and each edge in the state graph belongs to at least one test. When we execute these tests on the real system, we will test all possible transitions between the states.

Now, let us state the problem formally and provide algorithms that solve it.

\begin{definition}[Test suite generation problem (TSG)]
You are given a directed multigraph $G=(V,E)$ with a single source $s$: all vertices of $G$ are reachable from $s$. The \emph{test suite generation problem} requires to find a set of paths $P$ starting from $s$ such that every edge $e \in E$ is covered by some path $p \in P$, i.e., $e \in p$.

\emph{Minimum TSG problem} requires to find a set of paths of the smallest total length.
\end{definition}

The TSG problem does not require any minimization; thus, it can be solved in almost any possible way. Below we provide two approaches with varying time complexities. Also, we provide an algorithm for the minimum TSG problem.


Before explaining algorithms, we need to provide a definition. The diameter $D$ of a directed multigraph $G$ with source $s$ is equal to $\max_{v \in V} \delta\left(s, v\right)$, where $\delta\left(s, v\right)$ is the shortest distance from vertex $s$ to vertex $v$. In other words, it is equal to the maximum distance from source $s$ to any vertex in $V$. Note that in state graphs the diameter is usually very low, being $O(\mathrm{polylog}(|V|))$.


\subsection{Upper bound algorithm}
\label{upper-bound-algorithm}

Let us calculate the upper bound on the number of paths. The simplest upper bound will be if we use a separate path for each edge: $O(|E|)$ paths of total length $O(D \cdot |E|)$. Unfortunately, this upper bound is tight: consider a chain of length $D$ that has a star on its end~--- for each edge in the star, we need a separate path of length $D+1$.

The algorithm that matches this bound is pretty simple. Let us run BFS from $s$ and obtain a spanning tree of depth $D$. Then, for each edge $u \rightarrow v$, we take the path from $s$ to its source, $u$, appended with this edge.

This algorithm works in $O(D \cdot |E|)$ and creates $O(|E|)$ paths with total length $O(D \cdot |E|)$. Although the algorithm has an appealing time complexity, the state graph has a large number of edges, leading to an undesirable number of tests and their total length.

\subsection{Flow-based algorithms}
\label{efficient-graph-algorithm}

To reduce the test suite further, we need to devise an alternative approach. For that we need to remind of the circulation problem.

\begin{definition}[Circulation problem~\cite{tardos1985strongly}]
We are given a directed multigraph $G = (V, E)$ and two functions $l, u: E \rightarrow \mathbb{N} \, \cup \, \{0\}$. The circulation problem requires to find a flow function $f: E \rightarrow \mathbb{N} \, \cup \, \{0\}$ such that for each edge $l(e) \leq f(e) \leq u(e)$ and the incoming flow to each vertex $v$ is equal to the outcoming flow, $\sum\limits_{u \rightarrow v \in E} f(u \rightarrow v) = \sum\limits_{v \rightarrow w \in E} f(v \rightarrow w)$.

The minimum cost circulation problem additionally provides cost function $c: E \rightarrow \mathbb{N} \, \cup \, \{0\}$ and requires to find a flow function $f$ satisfying the requirements above with the minimal cost $\sum_{e \in E} c(e) \cdot f(e)$.
\end{definition}

\begin{remark}
The circulation problem can be solved using any algorithm for the maximum flow problem, for example, in $O(|V|^2 |E|)$ time using Dinic's algorithm~\cite{dinic1970algorithm}. While the minimum cost version can be solved in $O(|V||E| \log |V| \min(|V|C, |E| \log |V|))$~\cite{goldberg1989finding}, where $C$ is the maximum absolute value of the cost function.
\end{remark}

Now, we will reduce our TSG problem to the circulation problem. For that, we take the input to the TSG problem, a directed multigraph $G = (V, E)$, and add ``backward'' edges from all the vertices to the source vertex, $E'= \{ v \rightarrow s, \text{ for all } v \in V\}$. Then, on a new graph $G' = (V, E \cup E')$ we define two functions $l$ which is $1$ on edges from $E$ and $0$, otherwise, and $u$ which is always $+\infty$.



Suppose we found a solution to that circulation problem, represented by the flow $f$. Now we will get an answer to the TSG problem. For that, we create a new multigraph on the same set of vertices $V$ where each edge $e$ from $G'$ appears exactly $f(e)$ times. Note that due to the flow property, this new graph has an Eulerian circuit that is guaranteed to pass over ``original'' edges from $E$ at least one due to the requirement on $l(e) = 1$. Thus, the resulting test suite will be the paths obtained by splitting the found Eulerian circuit by ``backward'' edges from $E'$.
Note that, since the original graph has shortest paths of at most $D$, Dinic's algorithm~\cite{dinic1970algorithm} will work in $O(D \cdot |E| + D^2 \cdot |V|)$.


To solve the minimum TSG problem, in the reduction above, we can run an algorithm for the minimum cost circulation by choosing the cost function $c$ that is equal to $1$ for original edges in $E$ and $0$ for ``backward'' edges in $E'$: this will leave us with the minimum number of passes through original edges. The execution time of this algorithm will be $O(|V| |E| \log^2 |V|)$ by the remark above since the maximum absolute value of the cost function is $1$.

\begin{remark}
Although the algorithm for the usual TSG problem above may not yield the best solution, we use it to generate our test suite in the next section due to its low time complexity and good solutions in practice.
\end{remark}

\section{Practical results}
\label{practical-results-section}

We applied the described model-based testing approach to test the custom implementation of the distributed consensus algorithm Viewstamped Replication. This implementation of Viewstamped Replication features several major modifications, as outlined in the Introduction: incremental downloading of the log where nodes send only the missing parts; the ability to split an overloaded shard into multiple sub-shards; and the option to limit the number of replicas simultaneously running the garbage collection procedure.
%
%
All these modifications to the algorithm, along with Witness and Replica Set Reconfiguration capabilities of the original algorithm, have been verified using our model-based testing approach.

In Table~\ref{results-table}, we show how fast TLC generates the graph and how fast our test suite generation algorithms work on different upper bound values of the model. The benchmarks were executed on AMD Ryzen 5 4600H 3.00 GHz machine. As follows from the benchmark results, our system is capable of generating tens of thousands of tests per second when the graph is pre-generated. While the rate of testing implementation on these tests is determined mostly by the implementation itself (i.e., how fast can your actors' implementation process events), we report that we have achieved thousands of tests per second. 

\begin{table}
\caption{Results of several model-based test runs of Viewstamped Replication implementation}\label{results-table}
\centering
\begin{tabular}{|C{4.1cm}|C{1.7cm}|C{1.8cm}|C{1.8cm}|C{1.7cm}|C{1.9cm}|C{1.9cm}|}\hline
Values of constants & Diameter $D$ & Number of states $|V|$ & Number of edges $|E|$ & Path count $|P|$ & TLC graph generation & Test suite generation time \\\hline
$\texttt{ReplicasNumber} \leftarrow 3$ \newline $\texttt{MaxLogQueryCount} \leftarrow 3$ \newline $\texttt{MaxViewNumber} \leftarrow 2$ & $26$ & $207\,933$ & $1\,340\,753$ & $248\,626$ & 40 s & $6.17$ s \\\hline
$\texttt{ReplicasNumber} \leftarrow 3$ \newline $\texttt{MaxLogQueryCount} \leftarrow 4$ \newline $\texttt{MaxViewNumber} \leftarrow 2$ & $29$ & $2\,030\,058$ & $12\,921\,365$ & $2\,453\,461$ & 6 m & $55.23$ s \\\hline
$\texttt{ReplicasNumber} \leftarrow 3$ \newline $\texttt{MaxLogQueryCount} \leftarrow 5$ \newline $\texttt{MaxViewNumber} \leftarrow 1$ & $26$ &$581\,749$ & $3\,035\,901$ & $534\,711$ & 2 m & $13.72$ s \\\hline
$\texttt{ReplicasNumber} \leftarrow 5$ \newline $\texttt{MaxLogQueryCount} \leftarrow 1$ \newline $\texttt{MaxViewNumber} \leftarrow 2$ & $24$ & $166\,260$ & $2\,266\,623$ & $437\,158$ & 1 m & $9.86$ s \\\hline
$\texttt{ReplicasNumber} \leftarrow 5$ \newline $\texttt{MaxLogQueryCount} \leftarrow 2$ \newline $\texttt{MaxViewNumber} \leftarrow 2$ & $29$ & $5\,837\,512$ & $75\,127\,216$ & $15\,181\,992$ & 47 m & $265.58$ s \\\hline
\end{tabular}
\end{table}

\vspace{-0.3cm}
\section{Conclusion}
\label{conclusion-section}

In this work, we have designed an approach to verify that the implementation of a distributed system written in a commonly-used programming language corresponds to the model written in TLA+. Along the way we have designed a novel polynomial algorithm for generating a minimal exhaustive test suite from the model checking graph generated by the TLC model checker. Our approach was practically tested to verify the implementation of a consensus-based replication system used by VK.com. It was able to generate tens of thousands of tests per second, making it suitable for testing practical distributed systems.    

\newpage

\bibliography{references.bib}

\end{document}